\documentclass[prd,11pt]{revtex4}
\usepackage{amssymb,amsmath,amsthm,graphicx,psfrag}

\pdfoutput=1

\begin{document}

%%%%%%%%%%%%%%%%%%%%%%%%%%%%%%%%%%%%%%%%%%%%%%
%%%%%%%%%%%   New commands
%%%%%%%%%%%%%%%%%%%%%%%%%%%%%%%%%%%%%%%%%%%%%%

\def\be{\begin{equation}}
\def\ee{\end{equation}}
\def\bea{\begin{eqnarray}}
\def\eea{\end{eqnarray}}

\newcommand{\der}[2]{\frac{\partial{#1}}{\partial{#2}}}
\newcommand{\dder}[2]{\frac{\partial{}^2 #1}{{\partial{#2}}^2}}
\newcommand{\dderf}[3]{\frac{\partial{}^2 #1}{{\partial{#2} \partial{#3}}}}
\newcommand{\eq}[1]{Eq.~(\ref{eq:#1})}
\newcommand{\eqs}[1]{Eqs.~(\ref{eq:#1})}
\newcommand{\dd}{\mathrm{d}}
\newcommand{\PP}[4]{P^{#1\phantom{\underline{#2}}#3\phantom{\underline{#4}}}_{\phantom{#1}\underline{#2}\phantom{#3}\underline{#4}}}
\newcommand{\QQ}[3]{Q^{\phantom{\underline{#1}}#2\phantom{\underline{#3}}}_{\underline{#1}\phantom{#2}\underline{#3}}}

\title{Negative modes and the thermodynamics of Reissner-Nordstr\"om black holes}

\vskip 4cm

\author{R. Monteiro}
\email{R.J.F.Monteiro@damtp.cam.ac.uk}
\author{J. E. Santos}
\email{J.E.Santos@damtp.cam.ac.uk}

\vskip 0.2cm

\affiliation{DAMTP, Centre for Mathematical Sciences, University of Cambridge, Wilberforce Road, Cambridge CB3 0WA, UK}

\vskip 0.5cm

\date{\today}

\vskip 4cm

\begin{abstract}
We analyse the problem of negative modes of the Euclidean section of the Reissner-Nordstr\"om black hole in four dimensions. We find analytically that a negative mode disappears when the specific heat at constant charge becomes positive. The sector of perturbations analysed here is included in the canonical partition function of the magnetically charged black hole. The result obeys the usual rule that the partition function is only well-defined when there is local thermodynamical equilibrium. We point out the difficulty in quantising Einstein-Maxwell theory, where the so-called conformal factor problem is considerably more intricate. Our method, inspired by hep-th/0608001, allows us to decouple the divergent gauge volume and treat the metric perturbations sector in a gauge-invariant way.

\vskip 0.5cm
 
\end{abstract}

\maketitle

%\tableofcontents

%%%%%%%%%%%%%%%%%%%%%%%%%%%%%%%%%%%%%%%%%%%%%%%%%%%%%%%%%%%%%%%%%%%%%%%%%%%
%%%%%%%%%%%%%%%%%%%%%%%%%%%%%%%%%%%%%%%%%%%%%%%%%%%%%%%%%%%%%%%%%%%%%%%%%%%
\section{Introduction}

Shortly after the advent of black hole thermodynamics, the Euclidean path integral methods of quantum field theory at finite temperature were extended to semiclassical quantum gravity \cite{Hartle:1976tp,Gibbons:1976pt}. Gibbons and Hawking \cite{Gibbons:1976ue} proposed a construction for the partition functions of black holes. These were defined as path integrals with given boundary conditions, which correspond to fixing the temperature, through the periodicity in imaginary time, and possibly other quantities such as the electromagnetic charge for the Reissner-Nordstr\"om black hole. A black hole solution is seen as a saddle point of the path integral, its action being related to the thermodynamical free energy.

The semiclassical approach to the path integral allows for more than that. It is possible to go beyond the instanton approximation, corresponding to the classical black hole, and analyse small thermal (quantum) fluctuations around it. This was made possible after a better understanding of the path integral for perturbations, through the works of Gibbons, Hawking and Perry \cite{Gibbons:1978ac,Gibbons:1978ji}. They found that conformal perturbations of the metric, which always decrease the Euclidean action and seem to render the path integral divergent, are an unphysical artifact that can be eliminated by choosing a suitable integration contour and by applying a standard gauge fixing procedure. However, this treatment excluded any matter fields and one may (correctly) suppose that such a decomposition of the perturbations would be more involved.

The first application of those methods was the work of Gross, Perry and Yaffe \cite{Gross:1982cv}. They found that the Schwarzschild instanton possesses a nonconformal radial negative mode in the path integral for perturbations, which renders the partition function ill-defined. This was expected because the Hawking temperature formula for the Schwarzschild black hole, $T=(8 \pi M)^{-1}$, corresponds to a negative specific heat, so that there is no local thermodynamical stability. The authors further interpreted the instability as the possibility of spontaneous nucleation of black holes in hot flat space. The physics of black hole nucleation was clarified when York \cite{PhysRevD.33.2092} considered the partition function with boundary conditions at finite radius. Such a cavity, with fixed temperature on the wall, allows for two black hole solutions. The smaller radius solution is unstable and tends to the usual Schwarzschild case when the radius is taken to infinity. The larger radius solution is stable and its nucleation is thermodynamically allowed, since its free energy is inferior to the one of the hot flat space in the cavity.

A nontrivial test of the correspondence between local thermodynamical stability and a well-defined gravitational partition function was given by Prestidge \cite{Prestidge:1999uq}. He analysed the Schwarzschild-AdS instanton, soon after the AdS/CFT correspondence was proposed \cite{Maldacena:1997re}, and found numerically that the negative mode disappears when the specific heat of the black hole becomes positive \cite{Hawking:1982dh}. The curvature of AdS simulates the finite cavity in the asymptotically flat case. The correspondence was put on a firmer footing when Reall \cite{Reall:2001ag} showed, for a certain class of black holes, that a negative specific heat implies the existence of a negative mode. The proof of the converse result, however, remains elusive. Work on Euclidean negative modes of black holes includes also higher-dimensional solutions \cite{Asnin:2007rw}, the Taub-NUT and Taub-bolt instantons with \cite{Warnick:2006ih} or without \cite{PhysRevD.28.2420} cosmological constant, and connections to Ricci-flow \cite{Headrick:2006ti,Holzegel:2007zz}. A related subject is the one of black strings or branes whose classical stability has a correspondence with the local thermodynamical stability \cite{Gubser:2000ec,Gubser:2000mm,Reall:2001ag,Hirayama:2001bi,Hubeny:2002xn,Hirayama:2002hn,Hirayama:2008pf}; see \cite{Harmark:2007md} for a review. For instance, the Euclidean negative mode found by Gross, Perry and Yaffe \cite{Gross:1982cv} for the Schwarzschild black hole is related to the Gregory-Laflamme instability of the black string \cite{Gregory:1993vy}.

We point out that no study of negative modes of black hole instantons was performed with the inclusion of matter. One might be worried with the fact that such a theory is not renormalisable at one loop, which is indeed the case of Einstein-Maxwell theory \cite{PhysRevD.10.401}, so that we cannot compute quantum corrections. However, we can take an effective field theory approach \cite{Donoghue:1994dn}, where Einstein-Maxwell theory is regarded as the low energy limit of an underlying fundamental theory. The effective theory is valid up to a cut-off scale, after which ultraviolet completion effects become important. As long as the energies of the fields involved are nowhere near that scale, the perturbative quantisation of the nonrenormalisable effective theory is meaningful. It may help recalling that Einstein-Maxwell theory corresponds to the bosonic sector of four-dimensional $\mathcal{N}=2$ supergravity, which can be embedded into string theory.

Matter has not been considered in the study of negative modes of instantons before because the standard decomposition of the perturbations \cite{Gibbons:1978ac,Gibbons:1978ji}, which singles out the unphysical conformal modes, was performed for Einstein theory only (with or without cosmological constant). However, studies have been performed for cosmological applications by Gratton, Lewis and Turok \cite{Gratton:2000fj,Gratton:2001gw}, who considered the coupling to a scalar field. They look for well-behaved gauge-invariant perturbations which allow for a decoupling in the action between the relevant modes and the unphysical ones, which are integrated out. See Kol \cite{Kol:2006ux} for a general and systematic discussion of the same basic idea, which he calls ``power of the action,'' following its initial application to the Schwarzschild black hole and black string \cite{Kol:2006ga}. This corresponds to the procedure we adopt in this work. There is a further complication though. The problem of radial perturbations requires two gauge conditions (as the traceless and transverse conditions in pure gravity) for the metric perturbations. The method is inconsistent because the radial ansatz for the perturbation of the action only has a radial diffeomorphism invariance (one gauge choice), and including time does not allow for the construction of gauge-invariant quantities. The work of Kol \cite{Kol:2006ga} on the Schwarzschild black hole gives a solution to this problem. Considering a lift to one higher dimension, along which there is translational invariance, it is possible to construct gauge-invariant quantities, i.e. perturbations which are invariant for infinitesimal diffeomorphisms along the radial direction and along the extra dimension. If the action for the zero modes (infinite wavelength) along the extra-dimension reproduces the lower-dimensional action for perturbations, then the higher-dimensional action can be decomposed and the long wavelength limit taken when a simple reduced action is available.

Using this Kaluza-Klein method, we are able to study the four-dimensional magnetic Reissner-Nordstr\"om black hole in a sector of perturbations corresponding to the canonical ensemble, \emph{i.e.} fixed charge $Q$. We find that one negative mode still exists if the charge is small compared with the mass $M$, as expected, but disappears for $|Q| \geq \sqrt{3}M/2$, exactly when the specific heat becomes positive. This supports the validity of the canonical partition function as defined in \cite{Gibbons:1976ue}.

We mentioned that Einstein-Maxwell theory had not been considered for the perturbative path integral quantisation on black hole backgrounds. However, let us make some comments. First, a claim on the main result in the paper has been made in \cite{Prestidge:2000}. The negligence of the particularities of Einstein-Maxwell theory, as opposed to pure gravity, led to mistakes that we discuss in Appendix~\ref{app:tim}. Second, Miyamoto and Kudoh \cite{Miyamoto:2006nd,Miyamoto:2007mh} have analysed the classical stability of magnetically charged branes and verified that they are stable when there is local thermodynamical stability. This result is related to the works \cite{Gubser:2000ec,Gubser:2000mm,Reall:2001ag} mentioned before. The issue we address here is whether the partition function at one loop, defined as a saddle point approximation to a Euclidean path integral, conforms to the usual thermodynamical stability criterion. It is also worth mentioning that Einstein-Maxwell theory in four dimensions cannot be the result of dimensionally reducing Einstein-Maxwell theory in five dimensions, since there is then an extra scalar field. The fact that the Reissner-Nordstr\"om black hole is not straightforwardly related to a string will force us to be careful in our Kaluza-Klein action method, so that we make sure the correct quantum theory is obtained.

This paper is organised as follows. In Section~\ref{sec:EM}, we discuss the problems with quantising Einstein-Maxwell theory, stressing the differences with the pure Einstein case, and recall the relation between the boundary conditions of the path integral and the corresponding thermodynamical ensemble. In Section~\ref{sec:KK}, we explain the Kaluza-Klein action method to analyse the second-order action. In Section~\ref{sec:app0}, we describe the application of the method to the magnetic Reissner-Nordstr\"om black hole. We start by presenting an appropriate lift to five dimensions. We then construct the gauge-invariant quantities and obtain the reduced action. We find that the action possesses a negative mode when the specific heat at constant charge is negative. Finally, in Section~\ref{sec:conc}, we present the conclusions.

%%%%%%%%%%%%%%%%%%%%%%%%%%%%%%%%%%%%%%%%%%%%%%%%%%%%%%%%%%%%%%%%%%%%%%%%%%%
%%%%%%%%%%%%%%%%%%%%%%%%%%%%%%%%%%%%%%%%%%%%%%%%%%%%%%%%%%%%%%%%%%%%%%%%%%%
\section{\label{sec:EM}The Einstein-Maxwell path integral}%Jorge%
The partition function of a system of gravity coupled to electromagnetism is given by a path integral,
\be
\label{pathintegral}
Z=\int \dd[g]\dd[A] e^{-I[g,A]},
\ee
constructed from the Euclidean action
\be
\label{euclidaction}
I[g,A]=-\int_{\mathcal{M}} \dd^d x \sqrt{g}(R-F_{ab}F^{ab})-2\int_{\partial \mathcal{M}} \dd^{d-1} x \sqrt{h} (K-K_0),
\ee
where the first term corresponds to the usual Einstein-Hilbert action and the second term is the Maxwell action. The third term is the Gibbons-Hawking-York boundary term \cite{Gibbons:1976ue,York:1972sj}, required if the configurations summed over have a prescribed induced metric on $\partial \mathcal{M}$. Here $K$ represents the trace of the extrinsic curvature on $\partial \mathcal{M}$ and $K_0$ is the trace of the extrinsic curvature of flat spacetime, matching the black hole metric at infinity, necessary to render the on-shell action of asymptotically flat black hole solutions finite. Also, $F_{ab}$ are the components of an exact two-form obtained from the gauge field potential $A$, with the standard formula $F=\dd A$.

In the absence of electromagnetic sources, the purely gravitational action can be made arbitrarily negative for conformal transformations that obey the boundary conditions of the path integral, i.e. by geometries included in the sum. This apparent divergence in the functional integration, called the conformal factor problem, is circumvented by choosing an appropriate complex integration contour \cite{Gibbons:1978ac}, at least at the one loop level of the semiclassical quantisation. Furthermore, it was later shown in \cite{Gibbons:1978ji} that an orthogonal decomposition of the metric perturbations into a trace, a longitudinal-traceless and a transverse-traceless part, complemented by an appropriate gauge choice to deal with the diffeomorphism invariance of the action, leads to the complete decoupling of these components in the second-order action. It is a key requirement that the gauge choice kills the interaction between the trace and the longitudinal traceless parts. The final result confirms the prescription of \cite{Gibbons:1978ac}, and the decoupled trace can be integrated by choosing a suitable complex contour. Moreover, the scalar parts of the partition function, comprising the contributions from the tracelike (conformal) perturbations and from the scalar parts of the vector modes, both in the longitudinal traceless part of the metric and in the ghost vectors, cancel. This shows that the apparent nonpositivity of the purely gravitational action, and the consequent divergence of the path integral, are fixed by projecting out this contribution. We would like to address this problem in the presence of electromagnetism.

\subsection{\label{sec:EM_DEC}The second-order action}%Jorge%
In this section, we will perturb the action about a saddle point $(\tilde{g},\tilde{A})$, which we define here as a nonsingular solution of the equations of motion
\begin{subequations}
\be
\tilde{R}_{ab}-\frac{\tilde{g}_{ab}}{2}\tilde{R}=2\left(\tilde{F}_{a}^{\phantom{a}c}\tilde{F}_{bc}-\frac{\tilde{g}_{ab}}{4}\tilde{F}^{pq}\tilde{F}_{pq}\right)
\ee
and
\be
\tilde{\nabla}_a\tilde{F}^{ab}=0.
\ee
\end{subequations}
Small perturbations about this solution, $(h_{ab},a_b)$, are treated as quantum fields living on the saddle point background,
\be
\begin{array}{c@{\hspace{1 cm}}c@{\hspace{1 cm}}c}
g_{ab}=\tilde{g}_{ab}+h_{ab} & \text{and} & A_b = \tilde{A}_b + a_b.
\end{array}
\ee
We then perturb the action to second order,
\be
I[g,A]=I[\tilde{g},\tilde{A}]+I_2[h,a;\tilde{g},\tilde{A}]+\mathcal{O}(h^3,h^2 a,h a^2, a^3;\tilde{g},\tilde{A}),
\ee
so that the partition function can be approximated by a saddle point functional integral
\be
Z\simeq e^{-I[\tilde{g},\tilde{A}]}\int \dd[h]\dd[a]e^{-I_2[h,a]}\equiv e^{-I[\tilde{g},\tilde{A}]} Z_{(2)}.
\ee
The first-order term is absent because the instanton solution satisfies the equations of motion. The second-order action is given by
\begin{multline}
I_2 [h,a]= \frac{1}{4}\int \dd^d x \sqrt{g}\Big\{h^{ab}\Big[\Delta_L h_{ab}+2 \nabla_a \nabla^c h_{b c}-2 g_{ab}\nabla^m\nabla^n h_{mn}+g_{ab}\Box h+4 F_a^{\phantom{a}m}F_b^{\phantom{b}n}h_{mn}\\
+\frac{2 F^{mn}F_{mn}}{d-2}\Big(h_{ab}-\frac{g_{ab}}{2}h\Big)\Big]+16 a_c\Big(F_{ab}\nabla^a h^{bc}+\nabla^a F^{bc}h_{ab}+F^{ac}\nabla^b h_{ab}-\frac{1}{2}F^{ac}\nabla_a h\Big)\\
-8\Big(a_b\Box a^b-a_a a_b R^{ab}+\nabla_a a^a \nabla_b a^b\Big)\Big\},
\label{eq:EM_DEC_1}
\end{multline}
where we have defined $h=g^{ab}h_{ab}$ and $\sim$ was dropped out in all zero-th order quantities for the sake of notation. $\Delta_L$ is the so-called Lichnerowicz operator defined as
\be
\Delta_L P_{ab}=-\Box P_{ab}-2 R_{ambn}P^{mn}+2 R_{c(a}P_{b)}^{\phantom{b}c}.
\ee
This action can be checked to be invariant under the following gauge transformations
\be
\left\{
\begin{array}{c}
h_{ab}\to h_{ab}
\\
a_a \to a_a+\nabla_a \chi
\end{array}
\right.,
\ee
and
\be
\left\{
\begin{array}{c}
h_{ab}\to h_{ab}+\nabla_a V_b+\nabla_b V_a
\\
a_a \to a_a+V^b F_{ba}
\end{array}
\right. .
\ee
The first set of transformations corresponds to the $U(1)$ invariance associated with the electromagnetic field, and the second to the invariance under diffeomorphisms, conveniently mixed with the previous symmetry. Both these transformations require a gauge fixing procedure, by a suitable introduction of Fadeev-Popov determinants associated with each symmetry. However, the choice of gauge is essential to identify the relevant perturbations that lower the action and have physical meaning. We thus postpone this discussion to the end of this section.

At this stage we would like to discuss a fundamental property of \eq{EM_DEC_1}. If we only focus on four-dimensional instantons that are spherically symmetric, then the metric perturbations can be expanded into odd and even perturbations \cite{regge}. Substituting this general expansion in \eq{EM_DEC_1}, in the background of the magnetic Reissner-Nordstr\"om solution, the cross term gives zero for any value of the vector field perturbation $a_b$. If we further assume that whatever gauge choice that makes this problem tractable in four dimensions does not involve the mixture between metric perturbations and gauge potential perturbations, then the metric sector completely decouples from the gauge potential sector. In the electric Reissner-Nordstr\"om case, this decoupling does not occur, and a complete analysis of the problem requires the inclusion of such a term.

We further decompose the metric perturbations into a traceless and a trace part:
\be
h_{ab}=\phi_{ab}+\frac{1}{d}g_{ab}h,
\ee
After this expansion, the second-order action can be written as
\begin{multline}
I_2[\phi_{ab},h,a_a]=\frac{1}{4}\int \dd^d x\sqrt{g}\Big\{\phi^{ab}\Big[\Delta_L \phi_{ab}+4\Big(F_a^{\phantom{a}m}F_b^{\phantom{b}n}-\frac{g_{ab}}{d}F^{mc}F^{n}_{\phantom{n}c}\Big)\phi_{mn}+\frac{2}{d-2}F^{mn}F_{mn}\phi_{ab}\Big]\\
+\frac{8}{d}\phi_{ab}F^{ac}F^b_{\phantom{b}c}h+\left(\frac{d^2-3 d+2}{d^2}\right)h\Box h-\frac{d-4}{d^2}F^{ab}F_{ab}h^2-2 \nabla_a \phi^a_{\phantom{a}c}\nabla_b \phi^{bc}+\frac{2(d-2)}{d}\nabla_a \phi^{ab}\nabla_b h\\
+16 a_c\Big[F_{ab}\nabla^a \phi^{bc}+\nabla^a F^{bc}\phi_{ab}+F^{ac}\nabla^b \phi_{ab}-\frac{d-4}{2d}F^{ac}\nabla_a h\Big]-8\Big(a_b\Box a^b-a_a a_b R^{ab}+\nabla_a a^a \nabla_b a^b\Big)\Big\}.
\label{eq:EM_DEC_2}
\end{multline}
The first term in the second line of \eq{EM_DEC_2} is the term that makes the study of Einstein-Maxwell instantons more involved. It couples the traceless part of the metric with the trace part, and as a result one cannot identify the trace as being responsible for the divergent modes problem. There are several ways of tackling this problem. One might try to pick a particular gauge choice for the metric perturbations. This is exactly what one does in the purely gravitational case to remove the last term in the second line, by choosing a gauge of the form
\be
C_a[h]=\nabla^b\left(h_{ab}-\frac{1}{\beta}g_{ab}h\right),
\label{eq:123}
\ee
where $\beta$ is a constant to be conveniently fixed \cite{Gibbons:1978ji}. However, the authors were not able to find such a gauge when electromagnetism is introduced without making a shift in the gauge potential $a_a$. This shift results in a new coupling between the trace and the vector potential, invalidating once more the pure gravity interpretation of the trace. Furthermore, this new choice of gauge requires a new decomposition of the metric perturbations instead of the usual longitudinal/transverse decomposition corresponding to the choice (\ref{eq:123}).

An alternative approach to the problem would be to shift the trace component of the metric by a term proportional to $\Box^{-1}(\phi_{ab}F^{ac}F^b_{\phantom{b}c})$. This choice removes the problematic term, but also makes the second-order action dependent on the inverse box operator, rendering any further computations unpractical.

A third method, based on \cite{Kol:2006ga}, is to attempt to solve the problem by using Kaluza-Klein techniques, and will be the one followed in this paper.

\subsection{\label{sec:EM_BC}Boundary conditions}%Ricardo%
The partition function is defined as the path integral (\ref{pathintegral}) with appropriate boundary conditions. These boundary conditions specify the thermodynamical quantities which are held fixed in the ensemble. Here, we recall the discussion in \cite{Hawking:1995ap} for the Reissner-Nordstr\"om case.

As usual, the 3-metric on the boundary $\partial {\mathcal M} = S^1 \times S^2_{\infty}$ fixes the temperature $T=\beta^{-1}$, where $\beta$ is the periodicity of imaginary time. The boundary condition on the electromagnetic field at infinity typically fixes either the charge $Q$ or the potential $\Phi =Q/{r_+}$, as we shall discuss. Imposing a periodicity $\beta$, the leading order approximation to the path integral is the Reissner-Nordstr\"om instanton,
\be
\dd s^2 = f(r)\dd \tau^2+\frac{{\dd r^2}}{f(r)}+r^2\dd \Omega^2,
\ee
where $f(r) = 1-2 M/r+Q^2/r^2$, $\dd \Omega^2=\dd\theta^2+\sin^2\theta\dd \phi^2$, and $M$ and $Q$ are the black hole mass and charge, respectively. Since an instanton solution is required to be nonsingular, the mass and the charge are fixed by the boundary data and the condition of regularity at $r=r_+=M +\sqrt{M^2 -Q^2}$, the location of the outer horizon in the Lorentzian solution and of the ``bolt'' in the Euclidean solution. That condition is the formula for the Hawking temperature:
\be
T = \frac{r_+ -M}{2 \pi r_+^2}.
\ee

However, the action of the instanton depends on the boundary terms which make the fixing of quantities on $\partial {\mathcal M}$ consistent. Suppose one fixes the potential $A$ at a boundary at very large $r=R$, where $R$ will be taken to infinity. In the magnetic case, this corresponds to specifying the charge
\be
Q= \frac{1}{4 \pi} \int_{S^2_{\infty}} F,
\ee
which is determined by integrating the magnetic field strength $F_{\theta \phi}$ over $S^2_{\infty}$, which in turn is determined by $A$ on the boundary alone. But, in the electric case, the charge is computed using the dual of the electric field strength,
\be
Q= \frac{i}{4 \pi} \int_{S^2_{\infty}} \star F,
\ee
i.e. it requires fixing $F_{\tau r}$ (the $i$ is due to the use of imaginary time). The canonical ensemble, for which the charge $Q$ is fixed, includes the configurations for which derivatives of $A$ normal to the boundary are fixed. The thermodynamical quantity associated with specifying $A$ on $\partial {\mathcal M}$, in the electric case, is the electric potential at infinity $\Phi = -i A_\tau = Q/{r_+}$ (the gauge choice $A= -i (Q/r-\Phi) d\tau$ ensures that $A$ is regular on the horizon $r=r_+$). Fixing the potential $\Phi$ corresponds to the grand-canonical ensemble. For the canonical ensemble, the action (\ref{euclidaction}), with $d=4$, must include a boundary term appropriate for the variational problem in question,
\be
- 4 \int_{\partial {\mathcal M}} \dd^3 x \sqrt{h} F^{ab} n_a A_b,
\ee
i.e. to the fixing of $F_{\tau r}$ and thus of the electric charge $Q$. This term ensures that the Helmholtz energies of the electric and magnetic black holes coincide \cite{Hawking:1995ap}:
\be
F = -T \ln{Z_{\mathrm{canonical}}(\beta,Q)}= M-T S,
\ee
where $S = \pi r_+^2$ is the black hole entropy.

In the present work, we look at perturbations about the instanton such that the metric is fixed at the boundary. In the magnetic case, this sector leaves the electromagnetic potential $A$ unperturbed, which can be done consistently for $SO(3)$ symmetric backgrounds, as mentioned in the previous section. The sector is thus included in the canonical ensemble.

%%%%%%%%%%%%%%%%%%%%%%%%%%%%%%%%%%%%%%%%%%%%%%%%%%%%%%%%%%%%%%%%%%%%%%%%%%%
%%%%%%%%%%%%%%%%%%%%%%%%%%%%%%%%%%%%%%%%%%%%%%%%%%%%%%%%%%%%%%%%%%%%%%%%%%%
\section{\label{sec:KK}The Kaluza-Klein action method}

We described in Section~\ref{sec:EM} the difficulties in quantising Einstein-Maxwell theory. The standard decomposition of the perturbations around the instanton \cite{Gibbons:1978ji} does not apply (see Appendix~\ref{app:tim}). However, a decomposition that explicitly decouples the divergent modes, the conformal modes in the case of pure gravity, is essential. We therefore look for a different approach.

In \cite{Kol:2006ga}, Kol addresses the problem of the negative mode of the Schwarzschild black hole by looking at the ``dynamical'' part of the action. This procedure, valid if the problem has a single nonhomogeneous dimension (the radial one here), was formalised in \cite{Kol:2006ux}. Instead of the treatment of \cite{Gross:1982cv}, which looks at the Lichnerowicz operator $\Delta_L$ acting on transverse-traceless metric perturbations, an auxiliary extra dimension $z$ is added. The extended space of metric perturbations and the dependence on the extra dimension allow for the construction of several gauge-invariant quantities. These decouple into two sectors, a ``dynamical'' part, which is the relevant reduced action, and a ``nondynamical'' part, which takes away the divergent modes. The action of the five-dimensional zero modes, i.e. the $k=0$ modes in a Fourier decomposition along $z$, or at least a sector of it, is the four-dimensional action. But, in four dimensions, the radial problem has one gauge transformation only, $\xi_r$, and the appropriate fixing of the gauge freedom requires a second condition (as in the traceless and transverse gauge). This is provided by the $\xi_z$ component if we use the auxiliary extra dimension.

A different way of looking at the extra dimension is to relate the black hole to a black string. In the Schwarzschild case, the (thermodynamical) Euclidean negative mode of the black hole \cite{Gross:1982cv},
\be
(\Delta_L h)^{4D}_{ab}= \lambda_\mathrm{GPY} h^{4D}_{ab},
\ee
corresponds to the (classical) Lorentzian Gregory-Laflamme instability \cite{Gregory:1993vy} of the black string,
\be
(\Delta_L h)^{5D}_{\hat{a}\hat{b}}=0 \quad \Rightarrow \quad (\Delta_L h)^{4D}_{ab}+ k_\mathrm{GL}^2 h^{4D}_{ab}=0.
\ee
The critical wavenumber is $k_\mathrm{GL}^2 = -\lambda_\mathrm{GPY}$. From the five-dimensional point of view, the Euclidean negative eigenvalue of the $k=0$ (off-shell) modes gives the critical wavenumber for (on-shell) instabilities along $z$ \footnote{A straightforward correspondence of this type will not hold in our case because of the nontrivial factor in front of the second term in (\ref{eq:nega_10}).}.

The difficulty in the charged black hole case is that the standard traceless-transverse gauge does not decouple the divergent modes. The Kaluza-Klein action method (``power of the action'', as Kol prefers) seems now the only one available. It requires:

\begin{itemize}
\item[(i)] A ``maximally general ansatz'' for the perturbation of the fields; i.e. the ansatz must reproduce all of the background field equations by variation of the metric, and must be closed for the relevant group of gauge transformations.
\item[(ii)] The lift must be such that the five-dimensional action for the $k=0$ perturbation modes is equivalent to the action for the perturbations around the four-dimensional black hole instanton. Actually, it suffices that a particular sector of the five-dimensional action satisfies this. We must then restrict ourselves to that sector of the path integral.
\end{itemize}

The lift we consider in this work, along a timelike direction $z$, is a ``magnetic string'' of a theory with electromagnetism and Chern-Simons term. We will need to restrict to a sector of the path integral by introducing a Delta functional on the space of perturbations. The restriction will ensure that we obtain the four-dimensional action when we look at $k=0$ modes.

Before we start, let us make two clarifications. First, how do we know that the divergent modes are being decoupled from the path integral? The decomposition of the path integral into ``dynamical'' and ``nondynamical'' parts leads to a `non-dynamical' action simply composed of squares of gauge-invariant quantities. One of them will have a minus sign, meaning that a rotation to the imaginary line is needed in order for the Gaussian path integral to converge. This is the exact analogy of the prescription of \cite{Gibbons:1978ac}. Second, the particular eigenvalue obtained from this method is not the same, in general, as the one obtained from the standard decomposition. The negative eigenvalue in \cite{Kol:2006ga} is quantitatively different from the one in \cite{Gross:1982cv} because the decomposition of the action is different. But the positivity properties of the action, i.e. whether a negative eigenvalue exists or not, must be the same.

%%%%%%%%%%%%%%%%%%%%%%%%%%%%%%%%%%%%%%%%%%%%%%%%%%%%%%%%%%%%%%%%%%%%%%%%%%%
%%%%%%%%%%%%%%%%%%%%%%%%%%%%%%%%%%%%%%%%%%%%%%%%%%%%%%%%%%%%%%%%%%%%%%%%%%%
\section{\label{sec:app0}Application to the magnetic Reissner-Nordstr\"om black hole}

\subsection{\label{sec:app} Lift to five dimensions}%Jorge%

In this section, we will apply the method described in Section~\ref{sec:KK} to the magnetic Reissner-Nordstr\"om black hole. The first nontrivial step is to find a five-dimensional system of gravity, possibly coupled to some fields, that reduces to four-dimensional Einstein-Maxwell upon a Kaluza-Klein reduction on a circle. This truncation was first discussed by the authors of \cite{LozanoTellechea:2002pn} in the context of Lorentzian signature. Here we straightforwardly extend it to Euclidean signature spacetimes.

We start with minimal five-dimensional supergravity with the action given by
\be
I^{(5)} = -\int \dd x^5 \sqrt{\hat{g}}\Big(\hat{R}-\hat{F}^{\hat{a}\hat{b}}\hat{F}_{\hat{a}\hat{b}}-\frac{2}{3\sqrt{3}}\frac{\hat{\varepsilon}^{\hat{a}_1\ldots\hat{a}_5}}{\sqrt{\hat{g}}}\hat{F}_{\hat{a}_1\hat{a}_2}\hat{F}_{\hat{a}_3\hat{a}_4}\hat{A}_{\hat{a}_5}\Big),
\label{eq:app_1}
\ee
where $\hat{ }$ represents quantities in five dimensions, $\hat{g}=|\det{g_{\hat{a}\hat{b}}}|$ and $\hat{F}=\dd \hat{A}$. Our dimensional reduction ansatz will take the generic form
\be
\dd s_5^2 = \Delta^{\gamma}\dd s_4^2+\epsilon \Delta (\dd z+2 \tilde{A}_a \dd x^a)^2,
\label{eq:app_2}
\ee
where $\gamma$ is a constant and $\Delta$ is related with the exponential of the dilaton field. Here $\epsilon=\pm1$, according to the signature of the five-dimensional spacetime. The signature of the four-dimensional space is chosen to be Euclidean. Also, $\partial_z$ is a Killing vector of the five-dimensional spacetime, meaning that $\Delta$, $g_{ab}$ and $\tilde{A}_a$ do not depend on $z$.

The reduction of the Maxwell field parallels that of the metric, and in particular we expand it as
\be
\hat{A}=A+\Sigma(\dd z+2 \tilde{A}),
\label{eq:app_3}
\ee
where both $A$ and $\tilde{A}$ only have components in the four-dimensional space, and again do not depend on $z$. Choosing $\gamma=-1/2$, setting $\Delta=\exp{(-4\phi/\sqrt{3}})$ and substituting in the action (\ref{eq:app_1}) yields the following form of the action:
\begin{multline}
I^{(5)}=-V \int \dd^4 x \sqrt{g}\Big[R-2\partial_a\phi \partial^a \phi-e^{-2\sqrt{3}\phi}\epsilon \tilde{F}^{ab}\tilde{F}_{ab}-e^{-\frac{2}{\sqrt{3}}\phi}H^{ab}H_{ab}-e^{\frac{4}{\sqrt{3}}\phi}\partial_a \Sigma \partial^a \Sigma \\
-\frac{2}{\sqrt{3}}\Sigma \frac{\varepsilon^{a_1\ldots a_4}}{\sqrt{g}}\big(H_{a_1a_2}-4 \tilde{A}_{a_1}\partial_{a_2}\Sigma\big)\big(H_{a_3a_4}-4 \tilde{A}_{a_3}\partial_{a_4}\Sigma\big)\Big],
\label{eq:app_4}
\end{multline}
where $V$ is the volume of the circle along which we are performing our dimensional reduction and $H = F+2\Sigma \tilde{F}$. Varying this action with respect to $g_{ab}$, $\tilde{A}_a$, $A_a$ $\phi$ and $\Sigma$ yields the following equations of motion
\begin{subequations}
\be
R_{ab}=2 \partial_a \phi \partial_b \phi+e^{\frac{4}{\sqrt{3}}\phi}\partial_a \Sigma \partial_b \Sigma+2 e^{-2 \sqrt{3}\phi}\varepsilon \Big(\tilde{F}_{ac}\tilde{F}_b^{\phantom{b}c}-\frac{g_{ab}}{4}\tilde{F}_{mn}\tilde{F}^{mn}\Big)+2 e^{-\frac{2}{\sqrt{3}}\phi}\Big(H_{ac}H_b^{\phantom{b}c}-\frac{g_{ab}}{4}H_{mn}H^{mn}\Big),
\label{eq:app_a_1}
\ee
\be
\epsilon \nabla_a(e^{-2 \sqrt{3}\phi}\tilde{F}^{ab})+2\nabla_a (\Sigma e^{-\frac{2}{\sqrt{3}}}H^{ab})+\frac{4}{\sqrt{3}}\frac{\varepsilon^{a_1 a_2 a_3 b}}{\sqrt{g}}H_{a_1 a_2} \Sigma \partial_{a_3}\Sigma = 0,
\label{eq:app_a_2}
\ee
\be
\nabla_a(e^{-\frac{2}{\sqrt{3}}\phi}H^{ab})+\frac{2}{\sqrt{3}}\frac{\varepsilon^{a_1 a_2 a_3 b}}{\sqrt{g}}H_{a_1 a_2} \partial_{a_3}\Sigma=0,
\label{eq:app_a_3}
\ee
\be
\Box \phi+\frac{\sqrt{3}}{2}\epsilon \tilde{F}_{ab}\tilde{F}^{ab}+\frac{1}{2 \sqrt{3}}e^{-\frac{2}{\sqrt{3}}\phi}H_{a b}H^{ab}-\frac{1}{\sqrt{3}}e^{\frac{4}{\sqrt{3}}\phi}\partial^a \Sigma \partial_a \Sigma =0,
\label{eq:app_a_4}
\ee
and
\begin{multline}
\nabla_a(e^{\frac{4}{\sqrt{3}}\phi}\nabla^a \Sigma)-\frac{1}{\sqrt{3}}\frac{\varepsilon^{a_1\ldots a_4}}{\sqrt{g}}\big(H_{a_1a_2}-4 \tilde{A}_{a_1}\partial_{a_2}\Sigma\big)\big(H_{a_3a_4}-4 \tilde{A}_{a_3}\partial_{a_4}\Sigma\big)\\
+4 \frac{\varepsilon^{a_1\ldots a_4}}{\sqrt{g}} H_{a_1 a_2}\tilde{A}_{a_3}\partial_{a_4}\Sigma-2e^{-\frac{2}{\sqrt{3}}\phi}H^{ab}\tilde{F}_{ab}=0.
\label{eq:app_a_5}
\end{multline}
\label{eq:app_5}
\end{subequations}
We now search for a consistent truncation in which the two fields $\phi$ and $\Sigma$ are set to zero. We are then left with two constraints on $\tilde{F}$ and $F$ coming from both Eqs. (\ref{eq:app_a_4}) and (\ref{eq:app_a_5})
\be
\begin{array}{c@{\hspace{1 cm}}c@{\hspace{1 cm}}c}
3 \epsilon \tilde{F}^{ab}\tilde{F}_{ab}+F^{ab}F_{ab}=0 & \text{and} &\star F+\sqrt{3}\tilde{F}=0,
\end{array}
\label{eq:app_6}
\ee
where $\star$ is the Hodge dual with respect to the four-dimensional geometry. If the four-dimensional manifold has Lorentzian signature, the second condition solves the first if $\epsilon =1$. However, if the four-dimensional manifold is assumed to have the Euclidean signature, one must require $\epsilon=-1$. We thus from now on choose $\epsilon=-1$. The equations of motion (\ref{eq:app_5}) then reduce to
\be
\begin{array}{cccc}
R_{ab}=-2\Big(\tilde{F}_{ac}\tilde{F}_b^{\phantom{b}c}-\frac{g_{ab}}{4}\tilde{F}_{mn}\tilde{F}^{mn}\Big)+2\Big(F_{ac}F_b^{\phantom{b}c}-\frac{g_{ab}}{4}F_{mn}F^{mn}\Big), & \nabla_a F^{ab}=0 & \text{ and } & \nabla_a \tilde{F}^{ab}=0.
\end{array}
\label{eq:neg_7}
\ee
subject to the second constraint in (\ref{eq:app_6}). In order to further simplify the equations of motion and to explicitly solve the constraint, we change from $F$ and $\tilde{F}$ to $F^{(1)}$ and $F^{(2)}$ given by
\be
\begin{array}{c@{\hspace{1 cm}}c@{\hspace{1 cm}}c}
F^{(1)}=\frac{1}{2}\star \tilde{F}-\frac{\sqrt{3}}{2}F & \text{and} & F^{(2)}=\frac{\sqrt{3}}{2}\star \tilde{F}+\frac{1}{2}F.
\end{array}
\ee
The constraint is solved by setting $\star F^{(2)}=0$, that is, $F^{(2)}=0$. Substituting everything in (\ref{eq:app_a_1}) yields
\be
\begin{array}{ccc}
R_{ab}=2\Big(F^{(1)}_{ac}{F^{(1)}}_b^{\phantom{b}c}-\frac{g_{ab}}{4}F^{(1)}_{mn}{F^{(1)}}^{mn}\Big) & \text{ and } & \nabla_a({F^{(1)}}^{ab})=0.
\end{array}
\ee
These are precisely the equations of motion that we were searching for, i.e. the Einstein-Maxwell equations in four dimensions. Note that both these equations can be deduced from the four-dimensional action
\be
I^{(4)} = -\int \dd x^4 \sqrt{g} \Big(R-F^{(1)}_{ab}{F^{(1)}}^{ab}\Big).
\ee
The four-dimensional quantities are related to the five-dimensional quantities via
\be
\begin{array}{c@{\hspace{1 cm}}c@{\hspace{1 cm}}c@{\hspace{1 cm}}c}
\dd s^2_5 =\dd s^2_4-(\dd z+2 \tilde{A}_a \dd x^a)^2, & \displaystyle{\tilde{F} = \frac{\star F^{(1)}}{2}} & \text{and} & \displaystyle{\hat{A} = -\frac{\sqrt{3}}{2}A^{(1)}.}
\end{array}
\label{eq:app_8}
\ee
We further remark that this truncation is only valid at the level of the equations of motion. In fact, substituting directly the ansatz (\ref{eq:app_8}) in \eq{app_4} gives
\be
-V \int \dd^4 x \sqrt{g}\Big(R -\frac{1}{2}F^{(1)}_{ab}{F^{(1)}}^{ab}\Big)\neq V I^{(4)}.
\label{eq:app_11}
\ee

We can now write the five-dimensional lift of both the magnetic and electric Reissner-Nordstr\"om instantons, respectively as
\be
\begin{array}{ccc}
\displaystyle{\dd s^2_5 = f(r)\dd \tau^2+\frac{{\dd r^2}}{f(r)}+r^2\dd \Omega^2-\Big(\dd z-\frac{Q}{r} \dd\tau \Big)^2},& \quad & \displaystyle{A = -\frac{\sqrt{3}}{2}Q \cos\theta \dd\phi}
\end{array}
\label{eq:app_9}
\ee
and
\be
\begin{array}{ccc}
\displaystyle{\dd s^2_5 = f(r)\dd \tau^2+\frac{{\dd r^2}}{f(r)}+r^2\dd \Omega^2-\Big(\dd z-i Q \cos\theta \dd\phi \Big)^2}, & \quad  & \displaystyle{A = -i\frac{\sqrt{3}Q}{2 r} \dd\tau},
\label{eq:app_10}
\end{array}
\ee
where $f(r) = 1-2 M/r+Q^2/r^2$, $\dd \Omega^2=\dd\theta^2+\sin^2\theta\dd \phi^2$, and $M$ and $Q$ are the black hole mass and charge, respectively. We are now ready to use the ``power of action''.

%%%%%%%%%%%%%%%%%%%%%%%%%%%%%%%%%%%%%%%%%%%%%%%%%%%%%%%%%%%%%%%%%%%%%%%%%%%
%%%%%%%%%%%%%%%%%%%%%%%%%%%%%%%%%%%%%%%%%%%%%%%%%%%%%%%%%%%%%%%%%%%%%%%%%%%

\subsection{Reducing the action}%Jorge%
The ``maximally general ansatz'' for the magnetic case can be written as
\be
\dd s^2_5 = e^{2 A(r,z)}[\dd\tau-\kappa(r,z)]^2+e^{2 B(r,z)}\dd r^2+e^{2 C(r,z)}\dd \Omega^2-e^{2\beta(r,z)}[\dd z-\Gamma(r,z)\dd \tau-\alpha(r,z)\dd r]^2,
\label{eq:nega_0}
\ee
and $A = -(\sqrt{3}Q/{2})\cos\theta \dd\phi$. With this ansatz, it is consistent, in terms of gauge invariance, not to perturb the electromagnetic potential. This conforms to what we have seen before, because in the magnetic case the metric perturbations decouple from the vector potential perturbations, and the former have a positive definite action. For the electric case, one would have to find a form of perturbing the solution (\ref{eq:app_10}), maintaining all variables involved only dependent on $r$ and $z$. However, the authors were not able to accomplish this. From now on, we will only consider the magnetic case, for which the metric is explicitly codimension one. We then perturb the quantities in \eq{nega_0} as
\be
\begin{array}{cc}
\displaystyle{A(r,z)=A_0(r)+a(r,z)}, & \displaystyle{B(r,z)=B_0(r)+b(r,z)},
\\
\\
\displaystyle{C(r,z)=C_0(r)+c(r,z)}, & \displaystyle{\Gamma(r,z) = \Gamma_0(r)+\gamma(r,z)}, 
\end{array}
\label{eq:nega_1}
\ee
where all lower case letters are perturbations, and thus absent in the background solution, which is given by
\be
\begin{array}{c@{\hspace{1 cm}}c@{\hspace{1 cm}}c@{\hspace{1 cm}}c}
\displaystyle{A_0=-B_0=\frac{1}{2} \log f}, & C_0(r)=\log r & \text{and} & \displaystyle{ \Gamma_0(r)=\frac{Q}{r}}.
\end{array}
\ee
At zero-th order, we exactly recover \eq{app_9}. We then substitute \eq{nega_1} into the action (\ref{eq:app_1}) and expand it to second order,
\be
I^{(5)}_2=\int \dd r \dd z (P^{a\phantom{I}b\phantom{J}}_{\phantom{a}I\phantom{b}J}\partial_a u^I\partial_b u^J+Q^{\phantom{I}a\phantom{J}}_{I\phantom{a}J}u^I\partial_a u^J+V_{IJ}u^Iu^J)
\label{eq:nega_2}
\ee
where $u=\{a,b,c,\alpha,\beta,\gamma,\kappa\}$, $I\in\{\underline{1},\ldots,\underline{7}\}$ and the independent nonvanishing components of the tensors $P^{a\phantom{I}b\phantom{J}}_{\phantom{a}I\phantom{b}J}$, $Q_{I\phantom{a}J}^{\phantom{I}a\phantom{J}}$ and $V_{IJ}$ are given in Appendix~\ref{app:1}. One can now expand all the fields in Fourier modes, take the limits $k=0$ and $\beta=0$, and compare the five-dimensional quadratic action with the four-dimensional counterpart. As we predicted in \eq{app_11}, the two actions are not equal. Instead, they differ by the following term:
\be
I^{(5)}_{2,k=0,\beta=0}-I^{(4)}_2 = -\int \dd r\frac{\{r^2\partial_r\gamma(r)+Q [a(r)+b(r)-2 c(r)]\}^2}{2 r^2}
\ee
Note that here we are perturbing the four-dimensional action only in the metric sector, and with an ansatz equal to the $l=0$ ansatz for the Schwarzschild perturbations used in \cite{Gross:1982cv}. We will see later how to deal with this difference in actions.

The second-order action (\ref{eq:nega_2}) is invariant under the following gauge transformations
\be
\begin{array}{l}
\delta a = e^{-2 B_0}\xi_r A_0'+e^{-2 A_0}\Gamma_0^2\partial_z\xi_z
\\
\delta b = e^{-2 B_0}(\partial_r \xi_r-B_0' \xi_r)
\\
\delta c = e^{-2 B_0}\xi_r C_0'
\\
\delta \alpha = \partial_z \xi_r+\partial_r \xi_z-e^{-2 A_0}\xi_z \Gamma_0 \Gamma_0'
\\
\delta \beta = -\partial_z \xi_z
\\
\delta \gamma = e^{-2 B_0}\xi_r \Gamma_0'+2 \Gamma_0 \partial_z \xi_z
\\
\delta \kappa = -e^{-2 A_0}[\xi_z\Gamma_0'+\Gamma_0(\partial_z \xi_r +\partial_r \xi_z -2 \xi_z A_0')],
\end{array}
\label{eq:nega_3}
\ee
where $'$ denotes differentiation with respect to $r$ in zeroth-order quantities. These correspond to infinitesimal diffeomorphisms along  the vector $\xi = \xi_r \dd r+\xi_z \dd z$.

The quadratic action (\ref{eq:nega_2}) can be cast in a different form, in which $\alpha$ and $\kappa$ only appear as $\partial_z \alpha$ and $\partial_z \kappa$, by integration by parts. This was expected because, in the $k=0$ sector, one does not need to set $\alpha$ or $\kappa$ to zero, as one does for $\beta$. This motivates the construction of the following gauge independent quantities
\be
\begin{array}{l}
\displaystyle{q^1 = a-\left(\frac{c A_0'}{T_0'}-e^{-2 A_0}\Gamma_0^2 \beta\right)}
\\
\displaystyle{q^2 = b+\left[\frac{c B_0'}{T_0'}-e^{-2 B_0}\partial_r\left(\frac{e^{2 B_0}c}{T_0'}\right)\right]}
\\
\displaystyle{q^3 = \partial_z \kappa-e^{-2 A_0}\left(\beta \Gamma_0'-2 \Gamma_0 \beta A_0'-\frac{e^{2 B_0}\Gamma_0}{T_0'}\partial^2_z c+\Gamma_0 \partial_r\beta\right)}
\\
\displaystyle{q^4 = \partial_z \alpha-e^{-2 A_0}\left[\beta \Gamma_0 \Gamma_0'+\frac{e^{2 (A_0+B_0)}}{T_0'}\partial^2_zc-e^{2 A_0}\partial_r \beta\right]},
\\
\displaystyle{q^5 = \gamma-\left(\frac{c \Gamma_0'}{T_0'}-2 \Gamma_0 \beta\right)}.
\label{eq:nega_4}
\end{array}
\ee
Plugging in the expressions for $u^I$, as a function of the $q^i$, in \eq{nega_2}, gives a quadratic action that only depends on the $q^i$s. As expected, the action can be entirely written in terms of gauge-invariant quantities. We now proceed to integrate the ``nondynamical'' quantities $q^2$, $q^3$ and $q^4$, which appear with no derivatives in the action, and can thus be integrated. The action for the perturbations can be written as
\be
I^{(5)}_2 = I^{(5)}_{2,q^1,q^5}+\int \dd r \dd z\left[q^{\tilde{i}}L_{\tilde{i}\tilde{j}}q^{\tilde{i}}-2 R_{\tilde{i}}q^{\tilde{i}}\right],
\label{eq:nega_5}
\ee
where $\tilde{i}\in\{2,3,4\}$, $L$ does not depend on $q^1$ or $q^5$ and $R$ depends on $q^1$, $q^5$ and their derivatives. In fact,
\be
L = \left[
\begin{array}{ccc}
\displaystyle{ \frac{3 Q^2}{2 r^2}-2} & \displaystyle{\frac{3 f Q}{2}} & \displaystyle{\frac{1}{2} (1+3 f) r}
\\
\\
\displaystyle{ \frac{3 f Q}{2}} & \displaystyle{-\frac{1}{2} f^2 r^2} & \displaystyle{-\frac{1}{2} f Q r}
\\
\\
\displaystyle{ \frac{1}{2} (1+3 f) r }& \displaystyle{-\frac{1}{2} f Q r} & \displaystyle{-\frac{Q^2}{2}}
\end{array}
\right],
\label{eq:nega_6}
\ee
\be
R=\left[
\begin{array}{c}
\displaystyle{\frac{1}{2} \left(\frac{q^1 Q^2}{r^2}-\frac{2 r \partial^2_z q^5 Q}{f}+\partial_r q^5 Q+2 r^2 \partial^2_z q^1-4 r f\partial_r q^1\right)}
\\
\\
\displaystyle{\frac{(Q^2-r^2) q^5+r f[r (q^5+r\partial_r q^5)-Q q^1]}{2 r}}
\\
\\
\displaystyle{\frac{1}{2} \left[\left(r-r f-\frac{2 Q^2}{r}\right) q^1+2 Q q^5+r (2 r f \partial_r q^1-Q \partial_r q^5)\right]}
\end{array}
\right]
\label{eq:nega_7}
\ee
and
\be
I^{(5)}_{2,q^1,q^5}=-\frac{(Q q^1+r^2 \partial_r q^5)^2}{2 r^2}.
\label{eq:nega_8}
\ee
We can thus integrate the $q^{\tilde{i}}$ by constructing the following shifted variables
\be
\tilde{q}^{\tilde{i}} = q^{\tilde{i}}-(L^{-1})^{\tilde{i}\tilde{j}}R_{\tilde{j}},
\label{eq:nega_9}
\ee
in terms of which the action (\ref{eq:nega_5}) can be rewritten as
\be
I^{(5)}_2=I^{(5)}_{2,q^1,q^5}-\int \dd r \dd z R_{\tilde{i}}(L^{-1})^{\tilde{i}\tilde{j}}R_{\tilde{j}}+I_{ND},
\ee
where
\be
I_{ND}=\int \dd r \dd z \tilde{q}^{\tilde{i}}L_{\tilde{i}\tilde{j}}\tilde{q}^{\tilde{i}}.
\ee
The integration over the $\tilde{q}^{\tilde{i}}$ is now trivial, and will not affect the $q^1$ and $q^5$ dependent part of the action. The new effective action for the variables $q^1$ and $q^5$ is
\be
I_{\mathrm{eff}}\equiv I^{(5)}_{2,q^1,q^5}-\int \dd r \dd z R_{\tilde{i}}(L^{-1})^{\tilde{i}\tilde{j}}R_{\tilde{j}}.
\ee
This last expression is too cumbersome to be shown. However, we still have freedom to rotate $q^1$ and $q^5$. If one sets
\be
\begin{array}{l}
q^1 = a_{11} q+a_{15} \tilde{q}^5
\\
q^5 = a_{51} q+a_{55}\tilde{q}^5
\end{array},
\ee
where
\be
\begin{array}{c@{\hspace{1 cm}}c@{\hspace{1 cm}}c@{\hspace{1 cm}}c@{\hspace{1 cm}}c}
\displaystyle{a_{11}=\frac{r^2+3 f r^2-5 Q^2}{2 \sqrt{6} f r^2}}, & \displaystyle{a_{15} = \frac{Q a_{55}}{r f}}, & \displaystyle{a_{51}=-\frac{Q}{\sqrt{6}r}} & \text{ and } & \displaystyle{\partial_r a_{55} = -\frac{Q^2a_{55}}{r^3}},
\end{array}
\ee
then $\tilde{q}^5$ completely decouples from the action, and the effective action reduces to the simple expression
\be
I_{\mathrm{eff}} = \int \dd r \dd z r^2\left[f (\partial_r q)^2-\left(1-\frac{Q^2}{r^2 f}\right)(\partial_z q)^2+V q^2\right],
\label{eq:nega_10}
\ee
where $V$ is given by
\be
V=-\frac{6 M^2}{r^2(3 M-2 r)^2}\left[1-\frac{4}{3}\left(\frac{Q}{M}\right)^2\right].
\ee

We are now in a position to explain why the Kaluza-Klein method works in this specific case. The path integral for perturbations in five dimensions $Z^5_{(2)}$ reduces to the one in four dimensions $Z^4_{(2)}$ if we only integrate over directions in which the difference between the two actions is zero, i.e. if we tune $\gamma$. So, one has a particular sector of $Z^5_{(2)}$,
\be
\tilde{Z}^5_{(2)} =\lim_{k=0} \int \dd [u^I]e^{-I^{(5)}_2[u^I]}\delta[Q(a+b-2c)+r^2\partial_r \gamma]\delta[\beta] = \tilde{Z}^4_{(2)},
\ee
where $\delta[.]$ should be regarded as a Dirac delta function in a functional space. Above, $\tilde{Z}^4_{(2)}$ denotes the restriction of the four-dimensional path integral to the sector of radial metric perturbations (we showed before that it is consistent, in the magnetic case, to leave the electromagnetic potential unperturbed). Introducing the gauge-invariant variables gives the following form
\begin{align}
\tilde{Z}^4_{(2)} & = \lim_{k=0} \int \dd [q^i]\dd[c]\dd[\beta]e^{-I^{(5)}_2 [q^i]}\delta\left[Q(q^1+q^2)+r^2\partial_r q^5-\frac{Q(Q^2-2 r^2 f)\beta}{r^2 f}+2 Q r \partial_r \beta \right]\delta[\beta] \nonumber\\
    & = \lim_{k=0} \int \dd [q^i]e^{-I^{(5)}_2 [q^i]}\delta[Q(q^1+q^2)+r^2\partial_r q^5] \nonumber\\
    & = \lim_{k=0} \int \dd [q]\dd[\tilde{q}^5]\dd[\tilde{q}^{\tilde{i}}]e^{-I_{\mathrm{eff}}[q]-I_{ND}[\tilde{q}^{\tilde{i}}]}\delta\left[\sqrt{\frac{2}{3}} Q \left(1-\frac{Q^2}{f r^2}+\frac{2 r^2}{r^2+3 f r^2-3 Q^2}\right)q+Q \tilde{q}^{\tilde{2}}+a_{55} r^2 \partial_r \tilde{q}^5\right]\nonumber \\
    & = \lim_{k=0} \int \dd [q]e^{-I_{\mathrm{eff}}[q]},
\end{align}
where each equality holds up to an infinite constant. It is the crucial fact that the final form of the action, $I_{\mathrm{eff}}[q]$, does not depend on $\tilde{q}^5$ that makes the two path integrals equivalent when we look at the $k=0$ sector. The coefficients $a_{ij}$ were chosen in such a way that the argument of the last functional Dirac delta depends on $\partial_r \tilde{q}^5$ and not on both $\tilde{q}^5$ and $\partial_r \tilde{q}^5$ simultaneously.

%%%%%%%%%%%%%%%%%%%%%%%%%%%%%%%%%%%%%%%%%%%%%%%%%%%%%%%%%%%%%%%%%%%%%%%%%%%
%%%%%%%%%%%%%%%%%%%%%%%%%%%%%%%%%%%%%%%%%%%%%%%%%%%%%%%%%%%%%%%%%%%%%%%%%%%

\subsection{The negative mode}%Jorge%
In this section, we will compute the negative mode of the action by directly studying the eigenvalue associated with \eq{nega_10}, when the $z$ dependence drops out ($k=0$). The action reduces to
\be
I_{\mathrm{eff},k=0} = \int \dd r r^2\left[f(\partial_r q)^2+V q^2\right].
\label{eq:nega_1_1}
\ee
We first remark that, if $|Q|/M\geq\sqrt{3}/2$, the potential $V$ becomes positive, and as a result the action does not have a negative mode beyond this value of the charge. This is in exact agreement with the thermodynamical prediction, as we shall see.

\begin{figure}
\centering
\includegraphics[width = 8 cm]{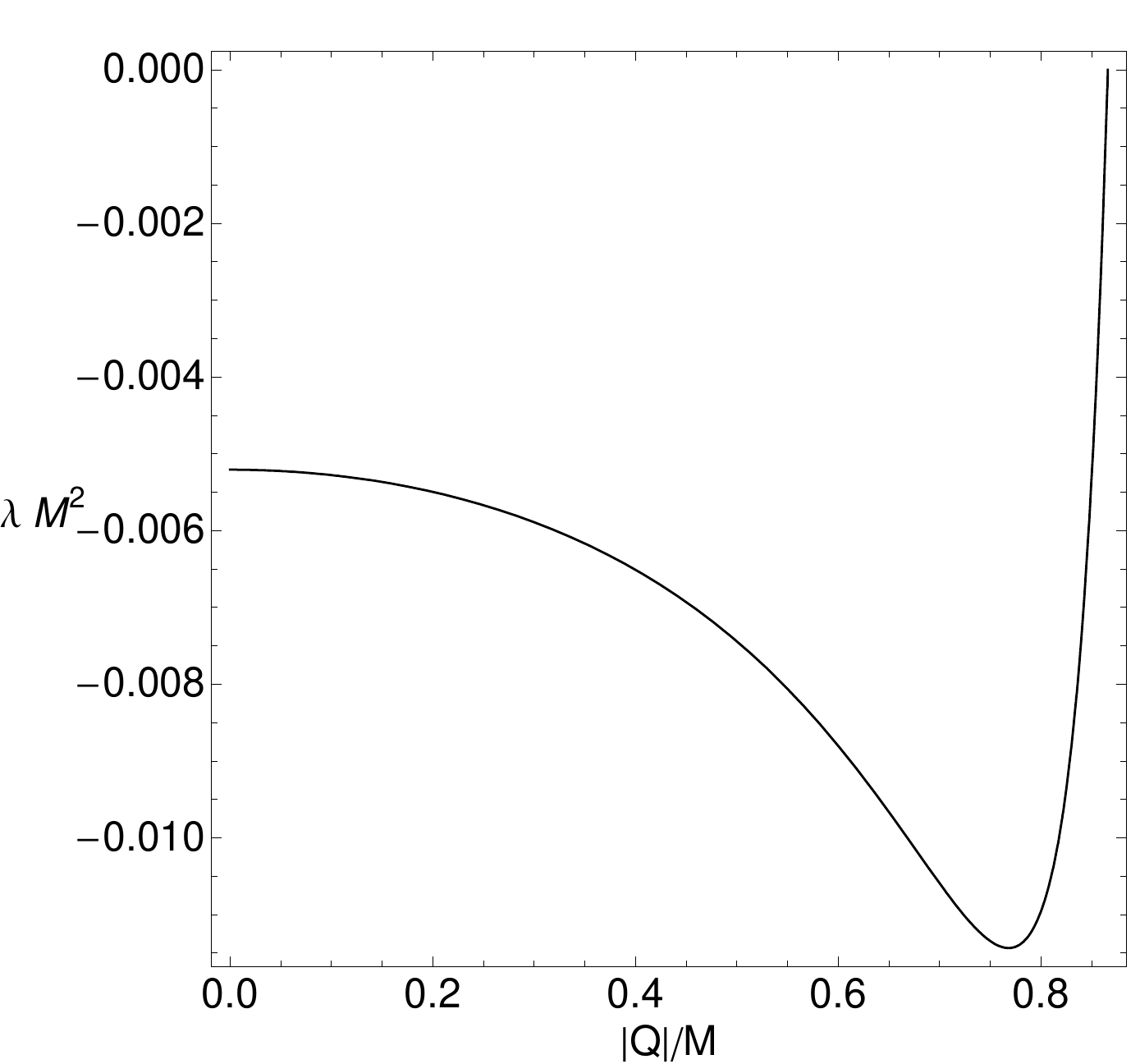}
\caption{\label{fig:nega_1}Evolution of the negative mode with $|Q|/M$.}
\end{figure}

To proceed, we change variables in \eq{nega_1_1}, in such a way that the resulting eigenvalue problem reduces to a one-dimensional Schr\"odinger equation. A convenient change of variables is
\be
\begin{array}{c@{\hspace{1 cm}}c@{\hspace{1 cm}}c}
\displaystyle{q=\frac{u}{\sqrt{\mathcal{T}}}} & \text{ and } & \displaystyle{y = -\frac{\mathcal{T}}{r_{_+}-r_{_-}}\log\left(\frac{r-r_{_+}}{r-r_{_-}}\right),}
\end{array}
\label{eq:nega_1_1_a}
\ee
where $r_{\pm}=M\pm\sqrt{M^2-Q^2}$ is the location of the black hole outer and inner horizons, respectively, and $\mathcal{T}$ is a constant chosen for later convenience. The action (\ref{eq:nega_1_1}) reduces to
\be
I_{\mathrm{eff},k=0}=\int \dd y \left[(\partial_y u)^2+\tilde{V} u^2\right],
\label{eq:nega_1_2}
\ee
where
\be
\tilde{V}=-\frac{2 \varepsilon (r_{_+}-r_{_-})^2 }{\mathcal{T}^2}\frac{e^{\frac{r_{_+}-r_{_-}}{\mathcal{T}}y}}{\left[1+\varepsilon e^{\frac{r_{_+}-r_{_-}}{\mathcal{T}}y}\right]^2}
\ee
and $\varepsilon=(r_{_+}-3 r_{_-})/(3 r_{_+}-r_{_-})$. Here we choose $\mathcal{T}=(r_{_+}-r_{_-})^2/\sqrt{\varepsilon}$, making the potential negative definite for all values of the black hole mass and charge. This choice is only valid for $\varepsilon>0$, that is, $|Q|/M<\sqrt{3}/2$. The eigenvalue problem can now be formulated as
\be
-\partial^2_{y} u+\tilde{V} u=\lambda u.
\label{eq:nega_1_4}
\ee
The change of coordinates (\ref{eq:nega_1_1_a}) maps $r=r_{_+}$ to $y=+\infty$ and $r=+\infty$ to $y=0$. As a result, the boundary conditions are now changed to $u$ being regular at $y=+\infty$, and integrable near $y=0$. Equation (\ref{eq:nega_1_4}) can be analytically solved for these boundary conditions and one finds the unique normalised bound state
\be
u=\frac{\sqrt{2}\varepsilon^{3/4}\sqrt{r_{_+}+r_{_-}}}{(r_{_+}-r_{_-})}\frac{e^{-\frac{(r_{_+}+r_{_-})y}{4 \mathcal{T}}}}{(\varepsilon-1)+(\varepsilon+1)\coth\left[\frac{(r_{_+}-r_{_-})y}{2 \mathcal{T}}\right]},
\ee
corresponding to
\be
-\lambda=\frac{2\sqrt{1-\eta^2}-1}{64 M^2 (1-\eta^2)^2(2\sqrt{1-\eta^2}+1)^2},
\label{eq:nega_1_5}
\ee
where $\eta=|Q|/M$. As expected, there is only one negative mode, and it disappears for $\eta\geq\sqrt{3}/2$ (Fig.~\ref{fig:nega_1}).

%%%%%%%%%%%%%%%%%%%%%%%%%%%%%%%%%%%%%%%%%%%%%%%%%%%%%%%%%%%%%%%%%%%%%%%%%%%
%%%%%%%%%%%%%%%%%%%%%%%%%%%%%%%%%%%%%%%%%%%%%%%%%%%%%%%%%%%%%%%%%%%%%%%%%%%

\subsection{Thermodynamical interpretation}%Ricardo%

Let us recall the thermodynamics of Reissner-Nordstr\"om black holes. The partition functions correspond to the different ensembles according to the boundary conditions of the path integral, as explained in Section \ref{sec:EM_BC}. In the magnetic case, fixing the temperature $T=\beta^{-1}$, by imposing the periodicity in imaginary time, and fixing the electromagnetic potential $A_a$ at infinity, which gives the magnetic charge $Q$, corresponds to the canonical ensemble.

The thermodynamical stability condition in the canonical ensemble is very simple: the specific heat at constant charge,
\be
C_Q = T \left( \der{S}{T} \right)_Q= - \frac{2 \pi r_+^2  (r_+ - r_-) }{r_+ - 3 r_-},
\ee
must be positive. This occurs for $\sqrt{3}M/2 < |Q| < M $, which is exactly the range we found previously for the disappearance of the negative mode in the partition function.

Had we studied the grand-canonical ensemble, we would expect a negative mode to persist. Here, the stability condition is the positivity of the so-called Weinhold metric \cite{Weinhold}, whose inverse is
\be
g_W^{\mu \nu} = -{\dderf{G}{y_\mu}{y_\nu}} = 
\left[
\begin{array}{cc}
\beta C_\Phi & \eta \\
\eta & \epsilon_T
\end{array}
\right],
\qquad y_\mu=(T,\Phi),
\ee
where $G = E - T S - \Phi Q$ is the Gibbs free energy, satisfying the first law of thermodynamics in the form
\be
\dd G=-S \dd T - Q \dd\Phi,
\ee
$\Phi = Q/r_+$ being the potential at infinity. The specific heat at constant potential is given by
\be
C_\Phi = T \left( \der{S}{T} \right)_\Phi = - 2 \pi r_+^2,
\ee
and the isothermal permittivity, really the capacitance here, is given by
\be
\epsilon_T = \left( {\der{Q}{\Phi}} \right)_T = \frac{T \eta^2}{C_\Phi - C_Q} = \frac{r_+ \left(r_+ - 3r_-\right)}{r_+ - r_-}.
\ee
The off-diagonal component is
\be
\eta = \left( {\der{Q}{T}} \right)_\Phi = \left( {\der{S}{\Phi}} \right)_T = -\frac{4 \pi  \sqrt{r_+ r_-} \,r_+^2}{r_+ - r_-}.
\ee
The second equality, given by the symmetry of the Hessian matrix, corresponds to the Maxwell relation. The matrix can easily be diagonalised by taking a new basis \mbox{$(\dd T,\dd \Phi) \to (\dd T,\dd \Phi + \epsilon_T^{-1} \eta \dd T)$},
\be
\left[
\begin{array}{cc}
\beta C_Q & 0 \\
0 & \epsilon_T
\end{array}
\right].
\ee
Since $C_Q$ and $\epsilon_T$ always have opposite signs, there is always one negative eigenvalue and the grand-canonical ensemble is unstable.

%%%%%%%%%%%%%%%%%%%%%%%%%%%%%%%%%%%%%%%%%%%%%%%%%%%%%%%%%%%%%%%%%%%%%%%%%%%
%%%%%%%%%%%%%%%%%%%%%%%%%%%%%%%%%%%%%%%%%%%%%%%%%%%%%%%%%%%%%%%%%%%%%%%%%%%
\section{\label{sec:conc}Conclusions}
In this paper, we have studied the problem of negative modes of the Euclidean section of the magnetic Reissner-Nordstr\"om black hole in four dimensions. Solving this problem within four dimensions seems very difficult, as the identification of which unphysical perturbations render the partition function divergent becomes considerably more intricate than in the vacuum case.

Following \cite{Kol:2006ga}, we devised a method to study this problem by lifting the magnetic Reissner-Nordstr\"om solution to five dimensions. The five-dimensional action is equal to the four-dimensional action, up to a quadratic term that can be set to zero by a suitable constraint, imposed by a functional Dirac delta, on the five-dimensional path integral. Furthermore, in five dimensions, the action can be solely written in terms of gauge-invariant variables, which in turn can be divided into two decoupled sectors: ``nondynamical'' and ``dynamical''. The former is algebraic, and can thus be readily integrated. The final form of the five-dimensional action depends on one gauge-invariant variable, and the study of its negative modes in the long wavelength limit, corresponding to four dimensions, is now possible.

We found complete agreement between the stability of the canonical ensemble and the existence of the negative mode. We analytically determined the eigenmode as a function of the black hole mass and magnetic charge, from which we concluded that the negative mode ceased to exist for $|Q|/M\geq\sqrt{3}/2$, which agrees with the thermodynamical prediction.

The Kaluza-Klein action method seems the only practical procedure to determine the quantum stability of gravity coupled to electromagnetism. However, a standard treatment of the metric and electromagnetic potential perturbations remains most desirable. The clarification of the divergent modes problem, analogous to the conformal factor problem of pure Einstein theory, would possibly lead to a better understanding of perturbative quantum gravity coupled to matter.

%%%%%%%%%%%%%%%%%%%%%%%%%%%%%%%%%%%%%%%%%%%%%%%%%%%%%%%%%%%%%%%%%%%%%%%%%%%
%%%%%%%%%%%%%%%%%%%%%%%%%%%%%%%%%%%%%%%%%%%%%%%%%%%%%%%%%%%%%%%%%%%%%%%%%%%
\section{\label{sec:thanks}Acknowledgments}

We are grateful to Malcolm Perry for suggesting this research topic, and for help and advice. We would also like to thank Stephen Hawking, Gustav Holzegel, Hari Kunduri and Claude Warnick for valuable discussions. This work was funded by Funda\c c\~ao para a Ci\^encia e Tecnologia (FCT, Portugal) through the grants SFRH/BD/22211/2005 (RM) and SFRH/BD/22058/2005 (JES).

%%%%%%%%%%%%%%%%%%%%%%%%%%%%%%%%%%%%%%%%%%%%%%%%%%%%%%%%%%%%%%%%%%%%%%%%%%%
%%%%%%%%%%%%%%%%%%%%%%%%%%%%%%%%%%%%%%%%%%%%%%%%%%%%%%%%%%%%%%%%%%%%%%%%%%%
\appendix
\section{\label{app:1}Independent components of the tensors $P$, $Q$ and $V$}
The independent components of $P^{a\phantom{I}b\phantom{J}}_{\phantom{a}I\phantom{b}J}$ are given by
\be
\begin{array}{l}
\begin{array}{cccc}
\PP{1}{1}{1}{3}=\PP{1}{3}{1}{3}=\PP{1}{3}{1}{5}=\PP{1}{3}{2}{4}=4 \PP{1}{6}{2}{7}, & \PP{1}{1}{1}{5}=\PP{1}{1}{2}{4}=2\PP{1}{6}{2}{7}, & \PP{1}{3}{2}{7}=\PP{1}{5}{2}{7}
\end{array}
\\
\begin{array}{cccccc}
\PP{1}{5}{2}{7}=2 \PP{2}{4}{2}{7}, & \PP{2}{1}{2}{3}=2\PP{2}{1}{2}{2}, & \PP{2}{2}{2}{3}=\PP{2}{3}{2}{3}, & \PP{2}{3}{2}{5}=2\PP{2}{5}{2}{5}, & \PP{2}{3}{2}{6}=2 \PP{2}{6}{2}{6}
\end{array}
\\
\begin{array}{cccccc}
\PP{2}{1}{2}{2}=-2 \PP{1}{6}{1}{6}, & \PP{1}{6}{2}{7}=-\frac{r^2 f}{2}, & \PP{1}{6}{1}{6}=-\frac{r^2}{2}, & \PP{1}{6}{2}{4}=\frac{Q r}{2}, & \PP{2}{3}{2}{3}=2r^2\left(1-\frac{Q^2}{r^2 f}\right) &
\end{array}
\\
\begin{array}{ccccc}
\PP{2}{2}{2}{5} = -\frac{Q^2}{f}, & \PP{2}{2}{2}{6}=-\frac{Q r}{f}, & \PP{2}{4}{2}{4}=-\frac{Q^2}{2}, & \PP{2}{4}{2}{7}=-\frac{Q r f}{2}, & \PP{2}{7}{2}{7}=-\frac{r^2 f^2}{2},
\end{array}
\end{array}
\ee
where all the other nonvanishing components can be obtained via the symmetry $P^{a\phantom{I}b\phantom{J}}_{\phantom{a}I\phantom{b}J}=P^{b\phantom{J}a\phantom{I}}_{\phantom{b}J\phantom{a}I}$. The terms that only involve one derivative were written using the auxiliary tensor $Q_{I\phantom{a}J}^{\phantom{I}a\phantom{J}}$, whose nonzero components are given by
\be
\begin{array}{l}
\begin{array}{cccc}
\QQ{1}{1}{1}=\QQ{5}{1}{1}=-2 \QQ{2}{1}{1}, & \QQ{1}{1}{3}=\QQ{5}{1}{3}=-2 \QQ{2}{1}{3}, & \QQ{1}{1}{5}=\QQ{5}{1}{5}=-2 \QQ{2}{1}{5}, & \QQ{5}{1}{6}=-3 \QQ{1}{1}{6}
\end{array}
\\
\begin{array}{ccccc}
\QQ{1}{1}{6}=\QQ{2}{1}{6}, & \QQ{3}{1}{1}=-2\QQ{2}{1}{1}, & \QQ{3}{1}{3}=-2\QQ{2}{1}{3}, & \QQ{3}{1}{5}=-2\QQ{2}{1}{5}, & \QQ{3}{1}{6}=-\QQ{6}{2}{4}=-2\QQ{2}{1}{6}
\end{array}
\\
\begin{array}{cccccc}
\QQ{6}{2}{4}=2 \QQ{2}{1}{6}, & \QQ{2}{2}{7}=3 \QQ{1}{2}{7}, & \QQ{1}{2}{4}=2\frac{Q^2}{r}-r+r f, & \QQ{1}{2}{7}= Q f, & \QQ{2}{1}{1}= 4 r f, & \QQ{2}{1}{6}=-Q
\end{array}
\\
\begin{array}{cccc}
\QQ{2}{1}{3}=-2\frac{Q^2}{r}+2 r+2 r f, & \QQ{2}{1}{5}=-\frac{Q^2}{r}+r+3 r f, & \QQ{2}{2}{4}=r(1+3 f), & \QQ{3}{2}{4}= -2\left(\frac{Q^2}{r}+2 r f\right)
\end{array}
\\
\begin{array}{ccccc}
\QQ{3}{2}{7}=2 Q-2 \frac{Q^3}{r^2}-4 Q f, & \QQ{5}{2}{4}=-\frac{3 Q^2}{r}, & \QQ{5}{2}{7}=Q \left(1-\frac{Q^2}{r^2}\right) & \text{ and } & \QQ{6}{2}{7}=-\frac{Q^2}{r}+r-r f.
\end{array}
\end{array}
\ee
Finally, the potential $V_{IJ}$ takes the following form:
\be
V_{IJ}=\left[
\begin{array}{ccccccc}
 \frac{3 Q^2-4 r^2}{2 r^2} & -\frac{Q^2}{2 r^2} & \frac{Q^2-2 r^2}{r^2} & 0 & \frac{5 Q^2-4 r^2}{2 r^2} & 0 & 0 \\
 -\frac{Q^2}{2 r^2} & \frac{3 Q^2-4 r^2}{2 r^2} & \frac{4 r^2-6 Q^2}{2 r^2} & 0 & \frac{Q^2}{2 r^2} & 0 & 0 \\
 \frac{Q^2-2 r^2}{r^2} & \frac{4 r^2-6 Q^2}{2 r^2} & \frac{6 Q^2-4 r^2}{r^2} & 0 & -\frac{Q^2+2 r^2}{r^2} & 0 & 0 \\
 0 & 0 & 0 & 0 & 0 & 0 & 0 \\
 \frac{5 Q^2-4 r^2}{2 r^2} & \frac{Q^2}{2 r^2} & -\frac{Q^2+2 r^2}{r^2} & 0 & -\frac{Q^2+4 r^2}{2 r^2} & 0 & 0 \\
 0 & 0 & 0 & 0 & 0 & 0 & 0 \\
 0 & 0 & 0 & 0 & 0 & 0 & 0
\end{array}
\right].
\ee

%%%%%%%%%%%%%%%%%%%%%%%%%%%%%%%%%%%%%%%%%%%%%%%%%%%%%%%%%%%%%%%%%%%%%%%%%%%
%%%%%%%%%%%%%%%%%%%%%%%%%%%%%%%%%%%%%%%%%%%%%%%%%%%%%%%%%%%%%%%%%%%%%%%%%%%
\section{\label{app:tim}Comment on a previous claim}

There is a previous claim in the literature \cite{Prestidge:2000} about our main result, namely that the negative mode in the partition function disappears for $|Q|/M \geq \sqrt{3}/2$. Since the work in question is not easily available and because the unfortunate mistake made there is instructive, we will make a comment on it.

If we ignore our initial discussion on the difficulty of quantising Einstein-Maxwell instantons, instead of purely gravitational ones, one may think that the use of the transverse-traceless (TT) gauge is still suitable. The change in the action caused by such perturbations, leaving the electromagnetic field unchanged, corresponds to the eigenvalue problem:
\be
\label{eigenTim}
({\mathcal O} h^{TT})_{ab} = \lambda {h^{TT}}_{ab},
\ee
where the perturbation operator is
\be
({\mathcal O} h^{TT})_{ab}= (\Delta_L h^{TT})_{ab} + F_{cd}F^{cd} h^{TT}_{ab}+4 F_a^{\phantom{a}c} F_b^{\phantom{b}d} h^{TT}_{cd}.
\ee
But this eigenvalue problem is ill-defined. The RHS of the eigenvalue equation (\ref{eigenTim}) is traceless-transverse, but not the LHS, due to the presence of the electromagnetic field.

In the case of \cite{Prestidge:2000}, the electric solution
\be
F = - i \frac{Q}{r^2}\dd\tau \wedge \dd r
\ee
is considered. An ansatz for the radial perturbations analogous to \cite{Gross:1982cv} (Schwarzschild) and \cite{Prestidge:1999uq} (Schwarzschild-AdS) is made. First, we note that leaving the electromagnetic field unperturbed is inconsistent, in the electric case, in terms of invariance for reparametrisations along $r$. Second, the eigenvalue equation (\ref{eigenTim}) has then four nontrivial components ($a=b$). One of them ($a=b=r$) is a second-order differential equation whereas the remaining components are third order equations. In both the Schwarzschild and the Schwarzschild-AdS cases, solving the second-order one solves the others. But not in the Reissner-Nordstr\"om case. The second-order equation actually has a negative mode that disappears for $|Q|/M \geq \sqrt{3}/2$, but the action is still not positive definite. The best way to see this is to use a probe traceless-transverse perturbation, a method explored in \cite{Rimajo} for rotating black holes. The readily available probe is proportional to the energy-momentum tensor of the electromagnetic field, which is obviously transverse and happens to be traceless in four dimensions. The result is that this probe perturbation always decreases the action, making clear the invalidity of the procedure above.

%%%%%%%%%%%%%%%%%%%%%%%%%%%%%%%%%%%%%%%%%%%%%%%
\bibliography{bibliography}

\end{document}